# Modification of the Coulomb potential from a Kaluza-Klein model with a Gauß-Bonnet term in the action

Harald H. Soleng*

*NORDITA, Blegdamsvej 17, DK-2100 Copenhagen Ø, Denmark*

Øyvind Grøn

*Oslo College of Engineering, Cort Adelers Gate 30, N-0254 Oslo, Norway*
*Institute of Physics, University of Oslo, P.O. Box 1048, N-0316 Blindern, Oslo 3, Norway*

(June 19, 1994; revised October 1, 1994)

## Abstract

In four dimensions a Gauß-Bonnet term in the action corresponds to a total derivative, and it does therefore not contribute to the classical equations of motion. For higher-dimensional geometries this term has the interesting property (which it shares with other dimensionally continued Euler densities) that when the action is varied with respect to the metric, it gives rise to a symmetric, covariantly conserved tensor of rank two which is a function of the metric and its first and second order derivatives. Here we review the unification of General Relativity and electromagnetism in the classical five-dimensional, restricted (with $g_{55} = 1$) Kaluza-Klein model. Then we discuss the modifications of the Einstein-Maxwell theory that results from adding the Gauß-Bonnet term in the action. The resulting four-dimensional theory describes a non-linear $U(1)$ gauge theory non-minimally coupled to gravity. For a point charge at rest we find a perturbative solution for large distances which gives a mass-dependent correction to the Coulomb potential. Near the source we find a power-law solution which seems to cure the short-distance divergency of the Coulomb potential. Possible ways to obtain an experimental upper limit to the coupling of the hypothetical Gauß-Bonnet term are also considered.

---

*Present address: Theory Division, CERN, CH-1211 Geneva 23, Switzerland
 Electronic address: Soleng@surya11.cern.ch

# I. INTRODUCTION

The geometrical unification of gravitation and electromagnetism obtained in the five-dimensional version of general relativity, constructed by Kaluza [1] and Klein [2], gave some beautiful results:

- Gravitation and electromagnetism is described within *one* theory.

- Electrical charge is explained as momentum of a neutral particle due to motion in the fifth dimension.

- The two signs of charge are understood as a consequence of two possible directions of motion around a closed fifth dimension.

- The electrical field of a charge is explained as the inertial dragging field due to the motion of a neutral particle in the $x^5$-direction [3,4].

- The quantization of the electric charge is found to be our four-dimensional description of the quantization of angular momentum, due to motion around the closed fifth 'cylinder-dimension' [5].

- The fine structure constant is given a geometrical interpretation, and it can be calculated from the ratio of two fundamental lengths—the radius of the fifth dimension and the Planck length [6].

- The theory can be generalized to include other forces by increasing the number of dimensions [7–9].

In spite of these results the Kaluza-Klein theories are not accepted as well-established physical theories—not even the five-dimensional one. This is mainly due to lack of new measureable consequences compared to ordinary Einstein-Maxwell theory. After all, the results mentioned have been obtained by introducing a fifth spatial dimension of space-time, and without new measureable predictions, it *is* difficult to believe in the existence of a closed fifth dimension with an estimated radius of $10^{-30}$ cm [5], making *direct* observations of this dimension virtually impossible.

Furthermore, the original Kaluza-Klein theory had a formal short-coming that made it less attractive. In order not to end up with a dilaton scalar field and an effective Brans-Dicke type scalar-tensor theory of gravitation, in disagreement with observations, one had to insert the condition $g_{55} = 1$ by hand. Doing this, the usual field equations of Einstein became an overdetermined system of equations, and the projection of the field equations into four-dimensional space-time resulted in the restriction $|\vec{E}| = |\vec{B}|$ for the magnitudes of the electrical and magnetic field strengths [10].

In 1973 this defect of the original theory was overcome by Leibowitz and Rosen [11]. They deduced the most general form of the field equations from the Einstein-Hilbert action under the constraint $g_{55} = 1$, and found a self-consistent system of equations. Using these equations to calculate the Kerr-like gravitational field of a point particle rotating in the fifth dimension, and projecting the solution into four-dimensional space-time, the Reissner-Nordström solution appeared [3].

This version of the five-dimensional Kaluza-Klein theory, which will be called "the restricted Kaluza-Klein theory," is mathematically self-consistent and in excellent agreement with observations. On the other hand, it could be argued that the restriction to $g_{55} = 1$ is aestetically unattractive and without a fundamental physical justification. It may be that the fundamental theory of Nature should contain a dilaton field, but experiments put very strong constraints on any scalar component in the gravitational field, and it seems that the dilaton, if it exists, must be very massive. From this perspective the restricted Kaluza-Klein model can be regarded as the effective theory when the dilaton is *very* heavy and effectively at rest in its potential well. Another objection to the restricted Kaluza-Klein model is that there is an exact agreement between the four-dimensional projection of the restricted Kaluza-Klein action and the action of the Einstein-Maxwell theory, and consequently there is no testable prediction of this model which is not also a prediction of the standard Einstein-Maxwell theory. Thus, it could be argued that the restricted Kaluza-Klein model is a mathematical curiosity rather than a testable scientific hypothesis.

In four-dimensional space-times the Einstein-Hilbert action is the unique geometrical action leading to second order differential equations for the metric. Einstein's field equations are therefore the only choice if one requires that the four-dimensional gravitational field equations are of second order. In Kaluza-Klein models there are other possibilities if one is willing to drop the restriction that the field equations are linear in second order derivatives of the metric. Lovelock [12] has shown that in general there are other symmetric, covariantly divergence free tensors of rank two that contain up to second order derivatives of the metric. These tensors are obtained by varying an action consisting of a sum of all dimensionally continued Euler densities for the relevant dimension. In five dimensions there are three possible tensors of this type—the metric tensor, the Einstein tensor, and the Lanczos[1] tensor. In the action these three tensors correspond to the cosmological constant (by convention the Euler density of zero dimensions), the Ricci scalar (the Euler density of two dimensions) and the Gauß-Bonnet term (the Euler density of four dimensions). Actions of this type give rise to Lovelock theories of gravity [12]. Such theories are ghost free [13,14] and parity conserving [15]. Thus, there is no physical reason to restrict the action to the Einstein-Hilbert term in Kaluza-Klein models. Results related to the development of string theories [16] also seem to imply a Gauß-Bonnet combination of curvature scalars as the lowest order correction to the Einstein-Hilbert term [13,14,17–26].

It is known that dimensional reduction of Kaluza-Klein theories with higher-order curvature terms lead to non-linearities of the gauge fields and to a non-minimal coupling to gravity [27–29]. These terms were first discovered by Horndeski [30] who determined all possible second-order four-dimensional vector-tensor field theories derivable from a variational principle and compatible with charge conservation and Maxwell's equations in flat space. Using a Kaluza-Klein model with a Gauß-Bonnet term, Buchdahl [27] found the same non-minimal couplings as Horndeski did. The total action of the dimensionally reduced five-dimensional Kaluza-Klein model with a Gauß-Bonnet term was first published by Müller-Hoissen [29].

---

[1] This tensor should not be confused with "Lanczos H-tensor" which is a rank three tensor that serves as a "potential" for the Weyl conformal tensor. Cornelius Lanczos was one of the pioneers in General Relativity and among the first to study actions with the Gauß-Bonnet term.

Over the last decade there have been much interest in the Gauß-Bonnet term and the corresponding corrections to Einstein's field equations. The wave propagation in the theory has been considered [31–33], and solutions corresponding to black holes [31,34–37] have been found and extended to the general Lovelock theory [15,38]. Some cosmological solutions are also known [34,39–49]. All the above solutions have been found using the *unrestricted* field equations, i.e. the scale factors of the additional dimensions have been regarded as functions of space and time.

In this paper we give a detailed discussion of the original five-dimensional *restricted* Kaluza-Klein theory and its generalization obtained by including a Gauß-Bonnet term in the action, and investigate the properties of the static solutions of a charged point particle within these theories. We do not select this theory because we believe that it is a realistic description of Nature, but because it is the simplest Kaluza-Klein model and because it contains standard Einstein-Maxwell theory. This model can therefore be used to probe the effect of the Gauß-Bonnet term and compare the results with standard physics. It is hoped that we in this way can capture some effects that could be present in more realistic models. We also keep in mind that the dimensionally reduced four-dimensional action could be regarded as fundamental. In this case the fifth dimension is a mathematical trick introduced in order to simplify the treatment of the rather complicated four-dimensional theory of a non-linear $U(1)$ gauge theory non-minimally coupled to gravity. To the best of our knowledge no solution of the *restricted* Kaluza-Klein theory with Gauß-Bonnet corrections have been published earlier.

## II. KALUZA-KLEIN THEORY

### A. The metric Ansatz

The basic idea of the original Kaluza-Klein unification of general relativity and electromagnetism is that space-time is five-dimensional and curved.[2] The new metric and connection components coming from the fifth dimension provide a geometrical description of the electromagnetic interaction. We are only able to observe four dimensions of space-time directly, and this fact must be accounted for by the model. Space-time is taken to be a five-dimensional manifold with Lorentzian signature. There is an isometry defined by a space-like Killing vector field, $\boldsymbol{\xi}$, so that

$$\xi_{(\mu;\nu)} = 0. \tag{1}$$

This means that space-time is homogeneous in the fifth direction, and there is no way to determine the position in the fifth dimension. The fifth dimension is also assumed to be compact. This means that in this direction the Universe is closed on a small scale. If the radius of the fifth dimension is small enough, it will be impossible to detect relative distances in this direction, and such a very compact cylindrical fifth dimension can not be observed directly. The radius of the fifth dimension is assumed to be constant. These two geometrical constraints—the isometry in the fifth direction and the constant radius—are called the first and second cylinder conditions.

---

[2]Some years earlier Nordström [50] had attempted the opposite way of unification when he unified his scalar theory of gravity with electromagnetism in a Maxwell theory in five-dimensional Minkowski space.

By using the Killing vector $\boldsymbol{\xi}$ as the fifth basis vector, the five-dimensional metric, $g_{\mu\nu}$, splits into the electromagnetic vector-potential $A_i$ and a space-time metric $^{(4)}g_{ij}$ in the following fashion[3]

$$g_{\mu\nu} = \begin{bmatrix} ^{(4)}g_{ij} + g^2 A_i A_j & g A_j \\ g A_i & 1 \end{bmatrix}. \tag{2}$$

In accordance with the first cylinder condition, all the fields are *independent* of the fifth coordinate. Furthermore, $g$ is a constant which will be determined later. We just note that since the fifth coordinate, $w$, has dimension length, the product $gA_j$ must be dimensionless. In units where $c = 1$ the vector potential $A_i$ has dimension $[A] = mass/charge$, the dimension of the constant $g$ must be $[g] = charge/mass$.

### B. Five-dimensional geodesics

Let us in accordance with the metric *Ansatz* (2), introduce a *fünfbein* defined by the an orthonormal one-form basis where the fifth component is the dual to $\boldsymbol{\xi}$. Such a basis is given by

$$\boldsymbol{\omega}^{\hat{\mu}} \in \{\boldsymbol{\omega}^{\hat{\imath}}, \quad \boldsymbol{\omega}^{\hat{5}} = \boldsymbol{d}w + g\boldsymbol{A}\} \quad \text{with} \quad \hat{\imath} \in \{\hat{1}, \hat{2}, \hat{3}, \hat{4}\}, \tag{3}$$

where $w$ is the fifth coordinate, $\boldsymbol{A} = A_{\hat{\imath}}\boldsymbol{\omega}^{\hat{\imath}}$, and $A_{\hat{5}} = 0$. There is no functional dependence on $w$ in any of the fünfbein coefficients, and the four first components do not involve $A_{\hat{\mu}}$ or $\boldsymbol{d}w$. Observe that a coordinate transformation $w \to w + \lambda$ is equivalent to a gauge transformation $\boldsymbol{A} \to \boldsymbol{A} + g^{-1}\boldsymbol{d}\lambda$. The reduction to the four-dimensional tetrad is obtained by ignoring the fifth component.

The structure coefficients are given by

$$\boldsymbol{d\omega}^{\hat{\mu}} = \frac{1}{2} C^{\hat{\mu}}{}_{\hat{\alpha}\hat{\beta}} \boldsymbol{\omega}^{\hat{\alpha}} \wedge \boldsymbol{\omega}^{\hat{\beta}}. \tag{4}$$

Note that with this definition $C^{\hat{\alpha}}{}_{\hat{\beta}\hat{\gamma}} = -C^{\hat{\alpha}}{}_{\hat{\gamma}\hat{\beta}}$. Consider now the structure coefficients coming from exterior differentiation of the four first one-forms of the basis (3). Since the basis forms $\boldsymbol{\omega}^{\hat{\imath}}$ do not depend upon $w$ and do not contain any terms proportional to $\boldsymbol{d}w$, and since $\{\boldsymbol{\omega}^{\hat{\imath}}\}$ constitutes the four-dimensional tetrad field, it follows that $C^{\hat{\imath}}{}_{\hat{\alpha}\hat{\beta}} = 0$ if $\hat{\alpha} = \hat{5}$ (or $\hat{\beta} = \hat{5}$), and that

$$d\boldsymbol{\omega}^{\hat{i}} = \frac{1}{2} C^{\hat{i}}{}_{\hat{\alpha}\hat{\beta}} \, \boldsymbol{\omega}^{\hat{\alpha}} \wedge \boldsymbol{\omega}^{\hat{\beta}} = \frac{1}{2} C^{\hat{i}}{}_{\hat{j}\hat{k}} \, \boldsymbol{\omega}^{\hat{j}} \wedge \boldsymbol{\omega}^{\hat{k}} = \frac{1}{2} \, {}^{(4)}C^{\hat{i}}{}_{\hat{j}\hat{k}} \, \boldsymbol{\omega}^{\hat{j}} \wedge \boldsymbol{\omega}^{\hat{k}} \quad (5)$$

where ${}^{(4)}C^{\hat{i}}{}_{\hat{j}\hat{k}}$ are the structure coefficients of the tetrad field. Exterior differentiation of the fifth basis one-form yields

$$d\boldsymbol{\omega}^{\hat{5}} = \frac{1}{2} C^{\hat{5}}{}_{\hat{\alpha}\hat{\beta}} \, \boldsymbol{\omega}^{\hat{\alpha}} \wedge \boldsymbol{\omega}^{\hat{\beta}} = \frac{1}{2} C^{\hat{5}}{}_{\hat{i}\hat{j}} \, \boldsymbol{\omega}^{\hat{i}} \wedge \boldsymbol{\omega}^{\hat{j}} = g \, d\boldsymbol{A}. \quad (6)$$

---

[3]Greek indices stand for five-dimensional components and Latin ones stand for the four-dimensional space-time. A hat over the index means that it refers to an orthonormal basis. We have used the metric signature +3.

From Eqs. (5) and (6) it follows that

$$C^{\hat{\alpha}}{}_{\hat{\beta}\hat{\gamma}} = 0 \quad \text{if} \quad \hat{\beta} = \hat{5} \quad \text{or} \quad \hat{\gamma} = \hat{5}. \quad (7)$$

The five-dimensional geodesic equation reads

$$\frac{dU^{\hat{\mu}}}{dS} + \Gamma^{\hat{\mu}}{}_{(\hat{\alpha}\hat{\beta})} U^{\hat{\alpha}} U^{\hat{\beta}} = 0 \quad (8)$$

where $S$ is the five-dimensional invariant path-length, and $U^{\hat{\mu}} = dx^{\hat{\mu}}/dS$ is the tangent vector field of the geodesic curve. The connection coefficients in an orthonormal basis are

$$\Gamma_{\hat{\alpha}\hat{\beta}\hat{\gamma}} = \frac{1}{2} \left( C_{\hat{\alpha}\hat{\beta}\hat{\gamma}} - C_{\hat{\gamma}\hat{\alpha}\hat{\beta}} - C_{\hat{\beta}\hat{\alpha}\hat{\gamma}} \right). \quad (9)$$

Together with Eq. (7) this gives

$$\Gamma^{\hat{5}}{}_{(\hat{\beta}\hat{\gamma})} = \Gamma_{\hat{5}(\hat{\beta}\hat{\gamma})} = -\frac{1}{2} \left( C_{\hat{\gamma}\hat{5}\hat{\beta}} + C_{\hat{\beta}\hat{5}\hat{\gamma}} \right) = 0. \quad (10)$$

Combining Eq. (10) with the $\hat{5}$-component of Eq. (8) we get $dU^{\hat{5}}/dS = 0$. There is therefore a constant of motion corresponding to the velocity component in the fifth direction,

$$U^{\hat{5}} = \beta \equiv \text{constant}. \quad (11)$$

The inner product of the tangent vector of the five-dimensional geodesic with itself is also a constant. Let this constant be given as $\epsilon \equiv -U^{\hat{\mu}} U_{\hat{\mu}}$ where $\epsilon$ is normalized to $-1$, $0$, or $+1$ for, respectively space-like, null, and timelike five-dimensional curves. This expression can be split into a four-dimensional part and a part from the fifth component of the tangent vector in the following manner: $U^{\hat{\mu}} U_{\hat{\mu}} = U^{\hat{k}} U_{\hat{k}} + U^{\hat{5}} U_{\hat{5}}$. The four-dimensional part is denoted by $\sigma^2 = -U^{\hat{k}} U_{\hat{k}}$. Combined with Eq. (11), this gives the relation

$$-\epsilon = \beta^2 - \sigma^2 \quad (12)$$

Since both $\epsilon$ and $\beta$ are constants, $\sigma$ is also a constant. All the three parameters $\epsilon$, $\beta$, and $\sigma$ are dimensionless (in units where $c = 1$).

A particle moving along a time-like curve in four-dimensional space-time has a four-velocity $u^i$ which satisifies the relation $u^k u_k = -1$, if we use units where $c = 1$. Since the corresponding five-dimensional quantities have $U^k U_k = -\sigma^2$, the four-velocity components $u^{\hat{i}}$ are related to the five-dimensional tangent vector components $U^{\hat{i}}$ by $u^{\hat{i}} \equiv U^{\hat{i}}/\sigma$. The scaling factor $\sigma$ has a simple physical interpretation. Since $u^i = dx^i/ds$ and $U^i = dx^i/dS$, where $s$ is the four-dimensional path-length, we get $\sigma = ds/dS$, i.e. $\sigma$ is the ratio of the four- and five-dimensional path lengths. By such a rescaling of the tangent vectors we can rewrite the four-dimensional components of the five-dimensional geodesic equation as

$$\frac{du^{\hat{i}}}{ds} + {}^{(4)}\Gamma^{\hat{i}}{}_{(\hat{j}\hat{k})} u^{\hat{j}} u^{\hat{k}} = -2\frac{\beta}{\sigma}\Gamma^{\hat{i}}{}_{(\hat{j}\hat{5})} u^{\hat{j}} \tag{13}$$

where we have used Eqs. (5), (9), and (10), implying that $\Gamma^{\hat{i}}{}_{\hat{j}\hat{k}} = {}^{(4)}\Gamma^{\hat{i}}{}_{\hat{j}\hat{k}}$, and Eqs. (7) and (9) leading to $\Gamma^{\hat{i}}{}_{\hat{5}\hat{5}} = 0$.

The terms on the left-hand side correspond to the four-acceleration, $a^{\hat{i}}$, and in the general theory of relativity, $a^{\hat{i}} = 0$ are the equations of motion of a freely falling particle. The terms on the right-hand side should be compared with the Lorentz force-law of electromagnetism. According to the Einstein-Maxwell theory, the equations of motion of a charged particle is

$$a^{\hat{i}} = \frac{e}{m} F^{\hat{i}\hat{j}} u_{\hat{j}} \tag{14}$$

where $F^{\hat{i}\hat{j}}$ are the components of the electromagnetic field tensor in the orthonormal basis.

From Eq. (9) we have

$$2\Gamma_{\hat{i}(\hat{j}\hat{5})} = -(C_{\hat{5}\hat{i}\hat{j}} + C_{\hat{j}\hat{i}\hat{5}}). \tag{15}$$

According to Eqs. (6) and (7)

$$C_{\hat{j}\hat{i}\hat{5}} = 0 \quad \text{and} \quad C_{\hat{5}\hat{i}\hat{j}} = 2g A_{[\hat{i};\hat{j}]} \tag{16}$$

where the covariant derivative comes from the exterior derivative of $\boldsymbol{A}$ in a non-coordinate basis. This gives

$$\Gamma_{\hat{i}(\hat{j}\hat{5})} = -\frac{1}{2} C_{\hat{5}\hat{i}\hat{j}} = -g A_{[\hat{i};\hat{j}]}. \tag{17}$$

With the identification of $F_{\hat{i}\hat{j}} \equiv 2A_{[\hat{i};\hat{j}]}$, it follows that $2\Gamma_{\hat{i}(\hat{j}\hat{5})} = -g F_{\hat{i}\hat{j}}$. As a consequence, the equation of motion (13) takes the form

$$a^{\hat{i}} = g\frac{\beta}{\sigma} F^{\hat{i}\hat{j}} u_{\hat{j}}. \tag{18}$$

This is Lorentz' force law provided the constants satisfy

$$e = gm\beta(\beta^2 + \epsilon)^{-1/2}, \tag{19}$$

where we have used Eq. (12) to eliminate $\sigma$. Thus, we have demonstrated that the equations of motion of charged particles in electromagnetic fields can be explained as a projection of five-dimensional geodesics down to four-dimensional space-time. This identification is independent of the five-dimensional field equations, as long as these field equations imply that the five-dimensional energy-momentum tensor is covariantly conserved.

### C. The action

#### 1. Einstein-Maxwell theory recovered

To assure that the field equations are compatible with the hypothesis of geodesic motion, one should start from an action principle or base the theory on covariantly divergence free tensors. The restricted Kaluza-Klein theory is based on the Einstein-Hilbert action

$$S_{\text{EH}} = \frac{1}{16\pi G_5} \int [R + \lambda(\xi^\mu \xi_\mu - 1)] \sqrt{|g|} d^5x \qquad (20)$$

where we have added a Lagrange multiplier $\lambda$ corresponding to the restriction $g_{55} = 1$. This term gives rise to the Leibowitz-Rosen [11] term in the five-dimensional field equations. Since it gives a vanishing contribution to the integral, it can be neglected when discussing dimensional reduction. $G_5$ is the five-dimensional gravitational constant. The five-dimensional Ricci scalar is

$$R = {}^{(4)}R - \frac{g^2}{4} F_{ij} F^{ij}. \qquad (21)$$

If we integrate out the fifth dimension, we get a factor of $2\pi R_5$ where $R_5$ is the radius of the fifth dimension. Since an overall constant factor is unimportant, we can identify

$$G_5 = 2\pi R_5 G \qquad (22)$$

where $G$ is Newton's constant. Then the dimensionally reduced Kaluza-Klein action is

$$S_{\text{KK}} = \frac{1}{16\pi G} \int {}^{(4)}R \sqrt{|{}^{(4)}g|} d^4x - \frac{g^2}{64\pi G} \int F_{ij} F^{ij} \sqrt{|{}^{(4)}g|} d^4x. \qquad (23)$$

This should be compared with the action of the Einstein-Maxwell theory. If we use the SI system of units

$$S_{\text{EM}} = \frac{1}{16\pi G} \int {}^{(4)}R \sqrt{|{}^{(4)}g|} d^4x - \frac{1}{4\mu_0} \int F_{ij} F^{ij} \sqrt{|{}^{(4)}g|} d^4x. \qquad (24)$$

Here $\mu_0$ is the magnetic permeability of vacuum. The dimensionally reduced action of the restricted Kaluza-Klein theory can be made identical to the Einstein-Maxwell action by a proper choice of $g$.

#### 2. Determination of g

As a result of the identification with the Einstein-Maxwell theory, we can identify the constant $g$ as

$$g^2 = 16\pi \epsilon_0 G \qquad (25)$$

where $\epsilon_0$ is the vacuum permittivity. In units with $c = 1$, it is related to $\mu_0$ by $\epsilon_0 = 1/\mu_0$. Using this result in Eq. (19), we find

$$e = 2(4\pi\epsilon_0 G)^{1/2} m\beta(\beta^2 + \epsilon)^{-1/2}. \qquad (26)$$

Thus, charge is a function of the fifth component $\beta$ of the tangent vector of the five-dimensional geodesic curve. Since a particle can move in two directions in the fifth dimension, charge can take either sign depending on the sign of $\beta$. If one requires that charged particles follow time-like trajectories in five-dimensional space-time ($\epsilon > 0$), one finds that all charged particles must have a mass of the order of the Planck mass or larger. This constraint is violated by huge factors by all observed charged particles. The electron, for example, is a factor of $10^{-22}$ lighter. Since tachyonic motion in five-dimensional space-time leads to no obvious conflict with four-dimensional observations [3], we do not let this distract us. To reduce the number of numerical and dimensional factors in the formulae we shall from now on use "geometrized units" of measurements [51]. In these units[4] $g = 2$.

### III. GAUß-BONNET MODIFIED KALUZA-KLEIN MODEL

#### A. Action and field equations

We shall consider the following action

$$S = \frac{1}{16\pi G_5} \int \left[ R + \ell^2 G_B + \lambda\left(\xi_\mu \xi^\mu - 1\right) \right] \sqrt{|g|} d^5 x. \qquad (27)$$

where $\xi^\mu$ is the Killing vector (1) and

$$G_B \equiv R_{\alpha\beta\gamma\delta} R^{\alpha\beta\gamma\delta} + R^2 - 4 R_{\alpha\beta} R^{\alpha\beta} \qquad (28)$$

is the Gauß-Bonnet term. $\lambda$ is a Lagrange multiplier introduced due to the second cylinder condition. The coupling constant $\ell$ has dimension of length. Variation of the action with respect to $\lambda$ and the metric gives the following field equations

$$\xi_\mu \xi^\mu - 1 = 0 \qquad (29)$$

and

$$R^\alpha{}_\beta - \frac{1}{2}\delta^\alpha{}_\beta \left[ R + \lambda(\xi^\mu \xi_\mu - 1) \right] + \ell^2 L^\alpha{}_\beta + \xi^\alpha \xi_\beta \lambda = 0. \qquad (30)$$

$L^\alpha{}_\beta$ is the Lanczos tensor

$$L^\alpha{}_\beta = 2 R^\alpha{}_{\gamma\delta\rho} R_\beta{}^{\gamma\delta\rho} + 4 R^{\gamma\delta} R^\alpha{}_{\gamma\delta\beta} + 2 R R^\alpha{}_\beta - 4 R^{\alpha\gamma} R_{\gamma\beta} - \frac{1}{2} \delta^\alpha{}_\beta G_B. \qquad (31)$$

It is a divergenceless counterpart to the Einstein tensor when the action is a Gauß-Bonnet term rather than the Einstein-Hilbert term.

---

[4] Since $g$ is dimensionless, charge is measured in units of length. The elementary charge is equal to the Planck length times the square root of the fine structure constant: $e = \sqrt{\alpha} \ell_P$. The Planck length is $\ell_P = 1.616 \times 10^{-33}$ cm.

Using Eq. (29) and taking the trace of Eq. (30), we determine the Lagrange multiplier:

$$\lambda = \frac{3}{2}R + \frac{\ell^2}{2}G_B. \tag{32}$$

Thus, the field equations are

$$R^\alpha{}_\beta - \frac{1}{2}\delta^\alpha{}_\beta R + \ell^2 L^\alpha{}_\beta + \frac{1}{2}\xi^\alpha \xi_\beta \left(3R + \ell^2 G_B\right) = 0 \tag{33}$$

The term involving $\xi^\alpha$ in the field equations (33) was first introduced by Leibowitz and Rosen [11] in the context of the Einstein-Hilbert-Kaluza-Klein theory. For a five-dimensional geometry satisfying the cylinder conditions, this term is also divergenceless when a Gauß-Bonnet term is added to the action.

The field equations (33) correspond to fourteen coupled, non-linear, second order field equations of the metric. The fifteenth equation is not independent of the others because the trace equation of (33) is automatically satisfied. This identity corresponds to the constraint $\xi^\mu \xi_\mu = 1$.

The vacuum field equations (33) can be generalized to five-dimensional non-vacuum by adding a source term on the left-hand side [11]. Since the left-hand side is covariantly divergenceless, the five-dimensional energy-momentum tensor must also be covariantly divergenceless. It follows that the world lines of free test particles are five-dimensional geodesics.

### B. The four-dimensional action

Following the same method of dimensional reduction as in Section II C, one finds that the Gauß-Bonnet term leads to a lot of interesting new terms [27–29]. With a metric of the form (2), the five-dimensional Gauß-Bonnet term (28) is

$$^{(5)}G_B = {}^{(4)}G_B + g^2 \nabla_k \left[F^{ij;k}F_{ij} - 2F^{ij}{}_{;j}F_i{}^k\right] + \frac{3}{16}g^4\left[(F_{ij}F^{ij})^2 - 2F_i{}^l F_l{}^m F_m{}^n F_n{}^i\right]$$
$$- \frac{1}{2}g^2 \left[{}^{(4)}R_{ijkl}F^{ij}F^{kl} + {}^{(4)}RF_{ij}F^{ij} - 4\,{}^{(4)}R_{ij}F^{ki}F_k{}^j\right]. \tag{34}$$

The four-dimensional Gauß-Bonnet term and the divergence term both contribute with a total derivative in the action and can be ignored. The remaining terms describe non-linear corrections to electromagnetism and a non-minimal coupling to gravity.

In terms of the electric and magnetic fields the non-linear terms of (34) reduce to

$$(F_{ij}F^{ij})^2 - 2F_i{}^l F_l{}^m F_m{}^n F_n{}^i = -8\,(\vec{B} \cdot \vec{E})^2. \tag{35}$$

For plane waves $\vec{E} \cdot \vec{B} = 0$, and thus there are no non-linear modifications of wave propagation in vacuum.

The non-linear and non-minimal coupling terms deduced here should be contrasted with similar terms in QED induced by one-loop vacuum polarization effects. For low frequencies, these effects are taken into account by the Euler-Heisenberg [53] effective Lagrangian. It contains [52] both the term of Eq. (35) and $(\vec{E} \cdot \vec{E} - \vec{B} \cdot \vec{B})^2$. In geometrized units the QED coefficients are of the order of $\ell_{\text{QED }F^4} \sim 10^9$ cm, and they are therefore much larger than the $\ell$ expected from a Kaluza-Klein theory. Vacuum polarization is an effect in which the photon for part of the time exists as a virtual electron-positron pair. It has been argued that this effectively gives the photon a finite size and that it as a result should feel tidal forces [54]. The effective QED action which takes into account this effect, contains all the three non-minimal coupling terms of Eq. (34), but the ratios of the coefficients are different than in our case. The effective coupling constants have been found [54] to be of the order of $\ell_{\text{QED } RF^2} \sim 10^{-14}$ cm.

## IV. KALUZA-KLEIN SOLUTIONS FOR A CHARGED POINT PARTICLE

### A. The metric in five and four dimensions

Let the five-dimensional line-element be given as

$$dS^2 = L^2(r)e^{2a(r)}dr^2 + r^2 d\Omega^2 - L^{-2}(r)dt^2 + V^2(r)dt^2 + 2V(r)dtdw + dw^2 \tag{36}$$

where units are chosen so that $c = 1$. $w$ is the coordinate which represents the fifth direction in the Kaluza-Klein geometry and

$$d\Omega^2 \equiv d\theta^2 + \sin^2\theta d\phi^2 \tag{37}$$

represents the area on the unit sphere.

Using an orthonormal frame in which the metric takes the form

$$\eta_{ij} \equiv \text{diagonal}\,[1, 1, 1, -1, 1], \tag{38}$$

and where the fifth basis vector is equal to the Killing vector $\boldsymbol{\xi}$, the fünfbein can be written

$$\begin{aligned}
\boldsymbol{\sigma}^{\hat{1}} &= L(r)e^{a(r)}\boldsymbol{dr} \\
\boldsymbol{\sigma}^{\hat{2}} &= r\boldsymbol{d\theta} \\
\boldsymbol{\sigma}^{\hat{3}} &= r\sin\theta \boldsymbol{d\phi} \\
\boldsymbol{\sigma}^{\hat{4}} &= L^{-1}(r)\boldsymbol{dt} \\
\boldsymbol{\sigma}^{\hat{5}} &= \boldsymbol{dw} + V(r)\boldsymbol{dt}
\end{aligned} \tag{39}$$

where $\boldsymbol{\sigma}^{\hat{5}}$ is the dual to the Killing vector $\boldsymbol{\xi}$. This basis is of the form specified in Eq. (3). Notice that we have chosen the numbering such that the time-direction corresponds to the fourth dimension, and the Kaluza-Klein ring is the fifth dimension. The notation here is chosen with care. If we split the five-dimensional expressions into the space-time metric and electromagnetic terms according to the prescription of Eq. (2), we find that the space-time metric is given by the four-dimensional line-element

$$ds^2 = L^2(r)e^{2a(r)}dr^2 + r^2 d\Omega^2 - L^{-2}(r)dt^2. \tag{40}$$

In the Einstein-Hilbert Kaluza-Klein model one finds [3] that $V(r) \sim q/r$, the Coulomb potential. According to the general results of Section II B, this interpretation does only depend on the geodesic motion of test particles and depends only on the generalized Einstein tensor being covariantly conserved. $V(r)$ is therefore proportional to the electromagnetic potential also in the Gauß-Bonnet extened Kaluza-Klein theory.

The line element (40), without further specification of the coefficients, represents the most general spherically symmetric, static space-time. Thus it has the symmetries that must be present in the space-time outside a charged point particle at rest.

### B. Exact solution of the restricted Kaluza-Klein equations with $\ell = 0$

Let us first look at the field equations without the Gauß-Bonnet term. As shown in Appendix C, it is in this case sufficient to consider the metric with $a(r) = 0$. Then Eq. (B5) becomes

$$\frac{2\,V'(r) + r\,V''(r)}{2\,r\,L(r)} = 0. \tag{41}$$

The solution is

$$V(r) = V_0 + \frac{a}{r} \tag{42}$$

which is of the form of a Coulomb potential plus a constant. This constant is the electrostatic potential energy at infinity, that is, something like a vacuum energy, but since the theory is invariant under changes of this constant (i.e., invariant under gauge transformations), this energy does not contribute to the gravitational field. According to Eq. (17) the electrostatic field is given by

$$E_{\hat{1}} = F_{\hat{1}\hat{4}} = g^{-1} C_{\hat{5}\hat{1}\hat{4}} = g^{-1} V', \tag{43}$$

where Eq. (A4) of Appendix A have been used. According to the Einstein-Maxwell theory $E_{\hat{1}} = -\varphi'(r)$, where $\varphi(r)$ is the Coulomb potential, $\varphi(r) = q/r$. The freedom of gauge transformations, allow us to determine the integration constants to be $V_0 = 0$ and $a = -2q$. As a result

$$V(r) = -\frac{2q}{r}. \tag{44}$$

By substitution of $V(r)$ and its derivatives into Eq. (B2), we get

$$\frac{q^2}{r^4} - \frac{1}{r^2} + \frac{1}{r^2\,L(r)^2} - \frac{2\,L'(r)}{r\,L(r)^3} = 0. \tag{45}$$

The solution is

$$L^{-2} = 1 - \frac{2m}{r} + \frac{q^2}{r^2} \tag{46}$$

where an integration constant has been identified with the mass, $m$, in order to give the correct Newtonian limit. The restricted Kaluza-Klein equations thus has a five-dimensional solution [3] which when projected to four dimensions reduces to the Reissner-Nordström metric and the Coulomb potential.

### C. Symmetries of the energy-momentum tensor in four dimensions

In the restricted Kaluza-Klein model with $\ell = 0$, a five-dimensional vacuum solution is a four-dimensional solution of the Einstein-Maxwell equations. The corresponding energy-momentum tensor in four-dimensional space-time is well known. It is, however, of interest to study the structure of the energy-momentum tensor in the four-dimensional picture in the case when $\ell \neq 0$. We interpret everything except the Einstein tensor as corresponding to some kind of energy-momentum in the four-dimensional picture (see Appendix F for details). This is reasonable, because the Lanczos tensor vanishes identically in four-dimensions.

For a space-time with the line-element (40) we get

$$^{(4)}G^{\hat{1}}{}_{\hat{1}} - {}^{(4)}G^{\hat{4}}{}_{\hat{4}} = \frac{2\,a'(r)}{e^{2\,a(r)}\,r\,L(r)^2}. \tag{47}$$

Hence, if $a'(r) = 0$, the radial component of the energy-momentum tensor is equal to the time-time component. For a fluid this would mean that $p_r = -\rho$ where $p_r$ is the radial pressure and $\rho$ is the energy density. In this case, observers measuring the $rr$ and $tt$ components of the energy-momentum tensor see a vacuum-like energy-momentum tensor with a tension equal to the energy density in the radial direction. Such energy-momentum tensors are boost invariant, and we call space-times of this type radially boost invariant.

In Eq. (F6) in Appendix F, the deviation from radial boost invariance is given by

$$\kappa T^{\hat{1}}{}_{\hat{1}} - \kappa T^{\hat{4}}{}_{\hat{4}} = \frac{2\,\ell^2\,V'(r)^2}{e^{4\,a(r)}\,r^2\,L(r)^2}$$

where $\kappa = 8\pi G$. If $\ell = 0$, that is, in the case of a Kaluza-Klein model without the Gauß-Bonnet term, the field equations (B2)–(B5) imply that $a'(r) = 0$, cf. Eq. (C5). If $a(r) = a_0$ is a constant, it can be absorbed by a redefinition of the coordinates, so in this case we may set $a(r) \equiv 0$. It follows that this solution of the simplest Kaluza-Klein theory is radially boost invariant in four dimensions.

Note that for a non-trivial potential $V(r)$, $a'(r) \neq 0$ if $\ell \neq 0$. Thus, the Gauß-Bonnet term explicitly breaks radial boost invariance of the electromagnetic energy-momentum tensor outside a charged point particle. As shown in Appendix F, the Gauß-Bonnet term also imply a violation of tracelessness of the four-dimensional energy-momentum tensor. This opens the possibility [55] that the non-minimal coupling allows for quantum creation of photons by the expansion of a homogeneous and isotropic universe, a process which is impossible for minimally coupled photons [56].

Due to the complexity of the field equations, we have put the expressions for the field equations and the solution method in the Appendices B and C.

### D. Correction to the Coulomb potential when $\ell \neq 0$

#### 1. Long distance

Even with the use of computer algebra programs, it seems impossible to find an exact solution to the field equations (B1). Let us therefore instead find the lowest-order corrections to the Coulomb potential induced by a hypothetical Gauß-Bonnet term. For that purpose we take a new look at Eq. (C5)

$$a'(r) = \frac{\ell^2 \, V'(r)^2}{e^{2a(r)} \, r}.$$

To find the lowest order correction to $a(r)$, we start with the Coulomb potential. If we use the Coulomb potential on the right-hand side, the equation takes the form

$$\left(\frac{e^{2a(r)}}{2}\right)' = -\left(\frac{\ell^2 \, q^2}{r^4}\right)', \tag{48}$$

from which we find

$$e^{a(r)} = 1 - \frac{\ell^2 \, q^2}{r^4}. \tag{49}$$

Let us now compute the corrections up to order $1/r^4$. All dimensionful parameters are absorbed in the fundamental length parameter $\ell$. Thus we introduce dimensionless parameters for charge and mass

$$q_1 \equiv \frac{q}{\ell} \quad \text{and} \quad m_1 \equiv \frac{m}{\ell}. \tag{50}$$

We also express the perturbation away from the Reissner-Nordström solution by dimensionless parameters, $v_2$, $v_3$, $v_4$, $c_3$ and $c_4$

$$e^{a(r)} \equiv 1 - \frac{q_1^2 \ell^4}{r^4} \tag{51}$$

$$V(r) \equiv -\frac{2 q_1 \ell}{r} + \frac{v_2 \ell^2}{r^2} + \frac{v_3 \ell^3}{r^3} + \frac{v_4 \ell^4}{r^4} \tag{52}$$

$$L^{-2}(r) \equiv 1 - \frac{2 m_1 \ell}{r} + \frac{q_1^2 \ell^2}{r^2} + \frac{c_3 \ell^3}{r^3} + \frac{c_4 \ell^4}{r^4} \tag{53}$$

where it is assumed that the dimensionless parameters are such that $p_i \ell^i \ll r^i$ (no sum over $i$). As we show in Appendix D, the field equations (33) to fourth order in $\ell$, lead to the following values for the dimensionless parameters in the above metric coefficients

$$v_2 = v_3 = c_3 = 0, \quad c_4 = -2 q_1^2 \quad \text{and} \quad v_4 = 4 m_1 q_1.$$

Hence, the perturbative solution for $a(r)$ is given by Eq. (49), whereas the effective Newtonian gravitational potential is given by $\Phi_N = -\frac{1}{2} - \frac{1}{2} {}^{(4)}g_{tt}$ in the space-time line-element (40)

$$\Phi(r)_N = -\frac{1}{2} + \frac{1}{2} L^{-2}(r) = -\frac{m}{r} + \frac{q^2}{2r^2} - \frac{\ell^2 q^2}{r^4} + \mathcal{O}(5). \tag{54}$$

The first term is the Newtonian potential from the central mass, and the second term can be understood as the contribution to the Newtonian potential from the electromagnetic field energy outside the mass. The last term is a correction from the Gauß-Bonnet term.

It is perhaps more interesting that the Coulomb potential, $\varphi(r)$, gets a mass-dependent long distance correction

$$\varphi(r)_\mathrm{L} = \frac{q}{r} - \frac{2\ell^2 mq}{r^4} + \mathcal{O}(5) \tag{55}$$

where $\varphi(r)$ is defined as the integral of the electric field $\varphi(r) \equiv -\int E_{\hat{1}} dr$ which by use of Eqs. (17) and (A4) leads to

$$\varphi(r) \equiv -\frac{1}{2}\int V'(r)e^{-a(r)}dr. \tag{56}$$

The reason why there is no $q^2$-correction to the Coulomb potential is that the $F^4$-terms in the action (34) do not contribute in the static, spherically symmetric model. Hence, if we neglect gravity, the standard Coulomb potential is recovered also in the non-linear theory.

### 2. Short distance

We have not been able to find the general solution to the field equations when $\ell \neq 0$, but the behaviour of the long-distance correction indicates that the electromagnetic interaction is weakened at short distances.

In Appendix E we have found a power-law solution (E4) valid in the limit where $r \ll \ell$ which supplements the other limit discussed above. The resulting electrostatic potential as defined in Eq. (56) is

$$\varphi(r)_\mathrm{S} = \varphi_0 \pm \frac{r}{2\ell}. \tag{57}$$

Thus the short distance effect of the Gauß-Bonnet term is equivalent to replacing the point charge with an effective extended core with a charge density $\sim 1/r$. The corresponding four-dimensional line-element is

$$ds^2 = \beta^2 dr^2 + r^2 d\Omega^2 - \frac{r^2}{\beta^2 r_0^2} dt^2. \tag{58}$$

Hence, the gravitational potential take the form of a harmonic oscillator potential.

By neglecting the mass-dependent correction, and making a simple matching of the two solutions (55) and (57) for the electrostatic potential at a radius $r = r_m$, we find

$$r_m = \sqrt{2q\ell} \quad \text{and} \quad \varphi_0 = \sqrt{\frac{2q}{\ell}}. \tag{59}$$

This gives the effective potential of Fig. 1. Note that this determination of $\varphi_0$ only is a matching condition for the two asymptotic solutions. It need not be the "physical" value. At this point another warning is appropriate. We have *not* proved that this interior solution really is part of the *same* solution as the exterior perturbative solution. Yet it is intriguing that the higher-order curvature terms seem to be able to regularize the divergence of the Coulomb potential. While this divergence is of no importance for atomic physics, because the nucleus is not a point charge, the difference is important for the electron (if this is a point particle) and for charged black holes. At the center of a charged black hole one cannot replace the point charge at the singularity with a finite-size charged object.

The metric (58) is compatible with a traceless effective energy-momentum tensor if $\beta^2 = 3$. In this case

$$\kappa T^{\hat{i}}{}_{\hat{j}} \sim \frac{1}{r^2} \text{diag}[0, 1, 1, -2]. \tag{60}$$

The fourth component is negative, in agreement with the strong energy-condition. This is the energy-momentum tensor of an ultra-relativistic fluid confined to the (2+1)-dimensional hyper-space normal to the radial direction.

In this connection it is worth mentioning that for any centrally symmetric potential, there is a one-to-one mapping of solutions of the non-relativistic radial equation of motion and solutions of (2+1)-dimensional Friedmann-Lemaître-Robertson-Walker (FLRW) cosmologies (see Appendix G for details). For the case of vanishing angular momentum, which is the case where a singularity threatens, the $1/r$ potential corresponds to a traceless $T_{\mu\nu}$ and Hooke's law to a (2+1)-dimensional $\Lambda$-term. These two potentials which in this way are singled out by the (2+1)-dimensional cosmology as the most symmetrical, represent the large $r$ and small $r$ solutions for the gravitational potential of the generalized Kaluza-Klein equations. The linear electrostatic potential found for small $r$ is less symmetric. We note, however, that its (2+1)-dimensional analogy is that of a string cloud cosmology. It is not evident if this only is a mathematical curiosity or if it has some deeper significance.

From the four-dimensional perspective, the $1/r^2$-dependence of the energy-momentum tensor implies that the integrated mass inside a sphere of radius $r$ is proportional to $r$ in this region. This means that there is no energy-divergence of a point charge. Again, it remains to be proven that this is the source of the perturbative exterior solution of the preceeding subsection.

### E. Experimental limit on $\ell$

In principle, the effect of the Gauß-Bonnet corrections on the hydrogen spectra can be used to put an upper limit on $\ell$. For an order of magnitude argument it is sufficient to consider a Bohr model, and we need only take into account the perturbative exterior solution because the exact behaviour of the potential near the nucleus makes very little difference in the electron wave function. The mass-dependent correction to the Coulomb potential changes the energy-difference between two Bohr orbits. The correction term changes the energy by

$$\Delta E_{GB} \sim \frac{8\ell^2 m e^2}{r^5} \Delta r \tag{61}$$

under a small radial displacement. This should be compared with the Coulomb potential for which

$$\Delta E \sim \frac{e^2}{r^2} \Delta r. \tag{62}$$

For a given $\Delta r$ between two Bohr orbits, this is equal to the energy $h\nu$ of a photon emitted when the electron makes a transition between the two states. The frequencies of atomic spectra are known with a certain uncertainty $\delta\nu$. It should be compared with the change due to the Gauß-Bonnet corrections

$$\frac{\Delta E_{GB}}{h\nu} \lesssim \frac{\delta\nu}{\nu}. \qquad (63)$$

The ratio of $\Delta E_{GB}$ and $h\nu$ at the Bohr radius, $r = a_0$, with $m = m_p$, the proton mass, is equal to $10^{-26}\ell^2$ cm$^{-2}$. This gives

$$\ell \lesssim 10^{13} \text{ cm } \sqrt{\frac{\delta\nu}{\nu}}. \qquad (64)$$

Since the spectra of the hydrogen atom are known with an accuracy of about one part in a million, this gives the bound

$$\ell \lesssim 10^{10} \text{ cm}. \qquad (65)$$

The QED value for the non-minimal coupling, $\ell_{QED\ RF^2}$, computed by Drummond and Hathrell [54] is more than twenty orders of magnitude below this limit. Comparision with the light deflection experiments give a bound of the same order of magnitude as in Eq. (65): the relative deviation from the Einstein value would be of the order of $(\ell/R_\odot)^2$ and since the radius of the sun is $R_\odot \approx 10^{10}$ cm, this could not improve the bound much. It appears therefore that the best bound on $\ell$ can be obtained from the non-linearities of the $F$-field rather than from the non-minimal coupling terms. For the non-linear terms, the QED value $\ell_{QED\ F^4}$ is around $10^9$ cm, and in order that the Kaluza-Klein non-linearties do not change the successful standard model predictions, e.g. the predictions for the anomalous magnetic moments, for which the standard theory agrees with the $g-2$ experiments with an accuracy of $10^{-11}$, $\ell$ has to be much less than this value.

There is in principle another possible test. As shown by Adler [57], the non-linear terms from vacuum polarization imply that electromagnetic waves propagating in an external magnetic field, will exhibit birefringence. That is, if $\vec{k}$ is the propagation direction, and $\vec{B}_0$ is the external field, then light linearly polarized in the $(\vec{k}, \vec{B}_0)$-plane has a different phase velocity than light with orthogonal polarization. The same effect is expected as a consequence of the non-linear term (35) of the present theory. The parallel propagation eigenmodes will have a phase velocity smaller than the orthogonal modes. Experimental possibilities for a direct test of this effect have been discussed in Ref. [58].

Neither of the above mentioned experimental tests can give a bound on $\ell$ much less than a cm, and this is many orders of magnitude above what one would expect from a Kaluza-Klein model where the compactified dimension has a radius of only $10^{-30}$ cm. Due to the weakness of gravity and the presence of non-linear terms in the effective QED Lagrangian which would shadow any non-linear Kaluza-Klein effects, it seems unlikely that any direct test of the Gauß-Bonnet terms can be made. Perhaps indirect evidence could be obtained from cosmology.

## V. CONCLUSION

We have shown that inclusion of a Gauß-Bonnet term in the action of the restricted Kaluza-Klein model leads to a non-linear $U(1)$ gauge theory non-minimally coupled to gravity. The ratios of the coupling constants of each of the new terms in the electromagnetic

Lagrangian are fixed by the theory. It has been shown that the theory gives a mass-dependent correction to the Coulomb potential at large distances and that it possibly leads to a regularization of the short-distance divergence. The mass-dependent correction is a genuine gravitational effect.

The theory contains a free parameter, $\ell$, of dimension length. In the limit $\ell \to 0$, the standard Einstein-Maxwell theory is recovered. The best upper bound on $\ell$ can be obtained by comparing the effect of the non-linearities of the Gauß-Bonnet corrected electromagnetism with the analogous effects due to vacuum polarization in QED.

Physical consequences of the non-minimal coupling terms are only significant for very strong gravitational fields. They could be of importance in the very early universe.

## ACKNOWLEDGMENTS

It is a pleasure to thank Paolo Di Vecchia for comments. We acknowledge use of the *Mathematica* package CARTAN in deriving the results of this paper. This work was supported in part by the Thomas Fearnley Foundation (H. H. S.), by Lise and Arnfinn Heje's Foundation, grant number 0F0377/1993 (H. H. S.).

## APPENDIX A: STRUCTURE AND CONNECTION COEFFICIENTS

The one-form basis (39) combined with Eq. (4) defines the following structure coefficients:

$$C^{\hat{2}}{}_{\hat{1}\hat{2}} = C^{\hat{3}}{}_{\hat{1}\hat{3}} = \frac{e^{-a(r)}}{rL(r)}, \tag{A1}$$

$$C^{\hat{3}}{}_{\hat{2}\hat{3}} = \frac{\cot\theta}{r}, \tag{A2}$$

$$C^{\hat{4}}{}_{\hat{1}\hat{4}} = -\frac{L'(r)e^{-a(r)}}{L(r)^2}, \tag{A3}$$

$$C^{\hat{5}}{}_{\hat{1}\hat{4}} = V'(r)e^{-a(r)}. \tag{A4}$$

The connection coefficients are

$$\Gamma_{\hat{1}\hat{2}\hat{2}} = \Gamma_{\hat{1}\hat{3}\hat{3}} = -\frac{e^{-a(r)}}{rL(r)}, \tag{A5}$$

$$\Gamma_{\hat{2}\hat{3}\hat{3}} = -\frac{\cot\theta}{r}, \tag{A6}$$

$$\Gamma_{\hat{1}\hat{4}\hat{4}} = -\frac{L'(r)e^{-a(r)}}{L(r)^2}, \tag{A7}$$

$$\Gamma_{\hat{1}\hat{4}\hat{5}} = \Gamma_{\hat{1}\hat{5}\hat{4}} = -\Gamma_{\hat{4}\hat{5}\hat{1}} = -\frac{1}{2}V'(r)e^{-a(r)}. \tag{A8}$$

## APPENDIX B: FIELD EQUATIONS

Let $C^\mu{}_\nu$ stand for the geometrical tensor on the left-hand side of the field equations (33). Symbolically, the field equations are given by

$$C^\mu{}_\nu = 0. \tag{B1}$$

Using the fünfbein frame (39), we find the following non-trivial components of the modified Kaluza-Klein equations

$$\begin{aligned}C^{\hat{1}}{}_{\hat{1}} &= \left[4\,e^{2\,a(r)}\,L(r) - 4\,e^{4\,a(r)}\,L(r)^3 - 8\,e^{2\,a(r)}\,r\,L'(r) - 12\,\ell^2 L(r)\,V'(r)^2 \right.\\ &\quad \left.+ 4\,\ell^2 e^{2\,a(r)}\,L(r)^3\,V'(r)^2 + e^{2\,a(r)}\,r^2\,L(r)^3\,V'(r)^2\right] \left[4\,e^{4\,a(r)}\,r^2\,L(r)^3\right]^{-1} \end{aligned} \tag{B2}$$

$$\begin{aligned}C^{\hat{2}}{}_{\hat{2}} &= \left[-4\,e^{2\,a(r)}\,L(r)^2\,a'(r) - 8\,e^{2\,a(r)}\,L(r)\,L'(r) + 4\,e^{2\,a(r)}\,r\,L(r)\,a'(r)\,L'(r)\right.\\ &\quad + 12\,e^{2\,a(r)}\,r\,L'(r)^2 - e^{2\,a(r)}\,r\,L(r)^4\,V'(r)^2 + 12\,\ell^2 L(r)^2\,a'(r)\,V'(r)^2 \\ &\quad \left.+ 8\,\ell^2 L(r)\,L'(r)\,V'(r)^2 - 4\,e^{2\,a(r)}\,r\,L(r)\,L''(r) - 8\,\ell^2 L(r)^2\,V'(r)\,V''(r)\right] \\ &\quad \left[4\,e^{4\,a(r)}\,r\,L(r)^4\right]^{-1} \end{aligned} \tag{B3}$$

$$\begin{aligned}C^{\hat{4}}{}_{\hat{4}} &= \left[4\,e^{2\,a(r)}\,L(r) - 4\,e^{4\,a(r)}\,L(r)^3 - 8\,e^{2\,a(r)}\,r\,L(r)\,a'(r) - 8\,e^{2\,a(r)}\,r\,L'(r) \right.\\ &\quad \left.- 4\,\ell^2 L(r)\,V'(r)^2 + 4\,\ell^2 e^{2\,a(r)}\,L(r)^3\,V'(r)^2 + e^{2\,a(r)}\,r^2\,L(r)^3\,V'(r)^2\right] \\ &\quad \left[4\,e^{4\,a(r)}\,r^2\,L(r)^3\right]^{-1} \end{aligned} \tag{B4}$$

$$\begin{aligned}C^{\hat{4}}{}_{\hat{5}} &= \left[2\,e^{2\,a(r)}\,r\,L(r)^3\,V'(r) + 12\,\ell^2 L(r)\,a'(r)\,V'(r) - 4\,\ell^2 e^{2\,a(r)}\,L(r)^3\,a'(r)\,V'(r) \right.\\ &\quad - e^{2\,a(r)}\,r^2\,L(r)^3\,a'(r)\,V'(r) + 8\,\ell^2 L'(r)\,V'(r) - 4\,\ell^2 L(r)\,V''(r) \\ &\quad \left.+ 4\,\ell^2 e^{2\,a(r)}\,L(r)^3\,V''(r) + e^{2\,a(r)}\,r^2\,L(r)^3\,V''(r)\right] \left[2\,e^{4\,a(r)}\,r^2\,L(r)^4\right]^{-1}. \end{aligned} \tag{B5}$$

In addition, the $C^{\hat{3}}{}_{\hat{3}}$ and $C^{\hat{5}}{}_{\hat{5}}$ components are non-trivial, but due to spherical symmetry $C^{\hat{3}}{}_{\hat{3}} = C^{\hat{2}}{}_{\hat{2}}$, and because of the trace identity of the field equations ($C^\mu{}_\mu = 0$), $C^{\hat{5}}{}_{\hat{5}}$ contains no new information.

## APPENDIX C: SOLUTION

Solving $C^{\hat{4}}{}_{\hat{5}} = 0$ with respect to $V''(r)$ gives

$$\begin{aligned}V''(r) &= -\left[2\,e^{2\,a(r)}\,r\,L(r)^3\,V'(r) + 12\,\ell^2 L(r)\,a'(r)\,V'(r) \right.\\ &\quad \left.- 4\,\ell^2 e^{2\,a(r)}\,L(r)^3\,a'(r)\,V'(r) - e^{2\,a(r)}\,r^2\,L(r)^3\,a'(r)\,V'(r) + 8\,\ell^2 L'(r)\,V'(r)\right] \\ &\quad \left[L(r)\left(-4\,\ell^2 + 4\,\ell^2 e^{2\,a(r)}\,L(r)^2 + e^{2\,a(r)}\,r^2\,L(r)^2\right)\right]^{-1}. \end{aligned} \tag{C1}$$

If we substitute this expression for $V''(r)$ back into $C^\mu{}_\nu$, and solve the resulting expression for $C^{\hat{2}}{}_{\hat{2}} = 0$ with respect to $L''(r)$, we get

$$\begin{aligned}
L''(r) = &-\Big[16\,\ell^2\,e^{2\,a(r)}\,L(r)^2\,a'(r) - 16\,\ell^2\,e^{4\,a(r)}\,L(r)^4\,a'(r) - 4\,e^{4\,a(r)}\,r^2\,L(r)^4\,a'(r) \\
&+ 32\,\ell^2\,e^{2\,a(r)}\,L(r)\,L'(r) - 32\,\ell^2\,e^{4\,a(r)}\,L(r)^3\,L'(r) - 8\,e^{4\,a(r)}\,r^2\,L(r)^3\,L'(r) \\
&- 16\,\ell^2\,e^{2\,a(r)}\,r\,L(r)\,a'(r)\,L'(r) + 16\,\ell^2\,e^{4\,a(r)}\,r\,L(r)^3\,a'(r)\,L'(r) \\
&+ 4\,e^{4\,a(r)}\,r^3\,L(r)^3\,a'(r)\,L'(r) - 48\,\ell^2\,e^{2\,a(r)}\,r\,L'(r)^2 + 48\,\ell^2\,e^{4\,a(r)}\,r\,L(r)^2\,L'(r)^2 \\
&+ 12\,e^{4\,a(r)}\,r^3\,L(r)^2\,L'(r)^2 + 20\,\ell^2\,e^{2\,a(r)}\,r\,L(r)^4\,V'(r)^2 - 4\,\ell^2\,e^{4\,a(r)}\,r\,L(r)^6\,V'(r)^2 \\
&- e^{4\,a(r)}\,r^3\,L(r)^6\,V'(r)^2 + 48\,\ell^4\,L(r)^2\,a'(r)\,V'(r)^2 + 16\,\ell^4\,e^{2\,a(r)}\,L(r)^4\,a'(r)\,V'(r)^2 \\
&+ 4\,\ell^2\,e^{2\,a(r)}\,r^2\,L(r)^4\,a'(r)\,V'(r)^2 + 32\,\ell^4\,L(r)\,L'(r)\,V'(r)^2 \\
&+ 32\,\ell^4\,e^{2\,a(r)}\,L(r)^3\,L'(r)\,V'(r)^2 + 8\,\ell^2\,e^{2\,a(r)}\,r^2\,L(r)^3\,L'(r)\,V'(r)^2\Big] \\
&\Big[4\,e^{2\,a(r)}\,r\,L(r)\,\big(4\,\ell^2 - 4\,\ell^2\,e^{2\,a(r)}\,L(r)^2 - e^{2\,a(r)}\,r^2\,L(r)^2\big)\Big]^{-1}. \quad \text{(C2)}
\end{aligned}$$

Substituting also this expression back into the expressions for $C^\mu_{\ \nu}$ and solving the resulting equation for $C^{\hat{1}}_{\ \hat{1}} = 0$ with respect to $L'(r)$, gives

$$\begin{aligned}
L'(r) = &\Big[L(r)\,\big(4\,e^{2\,a(r)} - 4\,e^{4\,a(r)}\,L(r)^2 - 12\,\ell^2\,V'(r)^2 + 4\,\ell^2\,e^{2\,a(r)}\,L(r)^2\,V'(r)^2 \\
&+ e^{2\,a(r)}\,r^2\,L(r)^2\,V'(r)^2\big)\Big]\Big[8\,e^{2\,a(r)}\,r\Big]^{-1}. \quad \text{(C3)}
\end{aligned}$$

Finally, substituting this expression back into the field equation components, we find that the only non-zero component is

$$C^{\hat{4}}_{\ \hat{4}} = \frac{2\,\big(\ell^2\,V'(r)^2 - e^{2\,a(r)}\,r\,a'(r)\big)}{e^{4\,a(r)}\,r^2\,L(r)^2}. \quad \text{(C4)}$$

The requirement $C^{\hat{4}}_{\ \hat{4}} = 0$ gives

$$a'(r) = \frac{\ell^2\,V'(r)^2}{e^{2\,a(r)}\,r}. \quad \text{(C5)}$$

Note that only the derivatives of $V(r)$ appear in the field equations. If $\ell \neq 0$, we can use Eq. (C5) to eliminate $V'(r)$ and $V''(r)$ from the field equations. In this way we find the following non-equivalent components:

$$C^{\hat{1}}_{\ \hat{1}} = -r^{-2} + \frac{1}{e^{2\,a(r)}\,r^2\,L(r)^2} + \frac{a'(r)}{r} + \frac{r\,a'(r)}{4\,\ell^2} - \frac{3\,a'(r)}{e^{2\,a(r)}\,r\,L(r)^2} - \frac{2\,L'(r)}{e^{2\,a(r)}\,r\,L(r)^3} \quad \text{(C6)}$$

$$\begin{aligned}
C^{\hat{2}}_{\ \hat{2}} = &-\frac{r\,a'(r)}{4\,\ell^2} - \frac{2\,a'(r)}{e^{2\,a(r)}\,r\,L(r)^2} + \frac{a'(r)^2}{e^{2\,a(r)}\,L(r)^2} - \frac{2\,L'(r)}{e^{2\,a(r)}\,r\,L(r)^3} + \frac{3\,a'(r)\,L'(r)}{e^{2\,a(r)}\,L(r)^3} \\
&+ \frac{3\,L'(r)^2}{e^{2\,a(r)}\,L(r)^4} - \frac{a''(r)}{e^{2\,a(r)}\,L(r)^2} - \frac{L''(r)}{e^{2\,a(r)}\,L(r)^3}
\end{aligned} \quad \text{(C7)}$$

$$\begin{aligned}
C^{\hat{4}}_{\ \hat{5}} = &\Big[-\frac{a'(r)}{e^{2\,a(r)}\,r\,L(r)^2} + \frac{5\,r\,a'(r)}{4\,\ell^2} + \frac{a'(r)}{r} + \frac{4\,a'(r)^2}{e^{2\,a(r)}\,L(r)^2} + \frac{4\,a'(r)\,L'(r)}{e^{2\,a(r)}\,L(r)^3} \\
&- \frac{a''(r)}{e^{2\,a(r)}\,L(r)^2} + a''(r) + \frac{r^2\,a''(r)}{4\,\ell^2}\Big]\Big[r\,V'(r)\,L(r)\Big]^{-1} \quad \text{(C8)}
\end{aligned}$$

# APPENDIX D: PERTURBATIVE SOLUTION

If we substitute the metric functions (51)–(53) into the expressions for the field equations (B1) and keep terms to fourth order in $\ell$, we get

$$C^{\hat{1}}{}_{\hat{1}} = \frac{\ell^3 \left(-3 c_4 \ell - 6 \ell q_1{}^2 - 2 c_3 r - 2 q_1 r v_2 + \ell v_2{}^2 - 3 \ell q_1 v_3\right)}{r^6} \tag{D1}$$

$$C^{\hat{2}}{}_{\hat{2}} = \frac{\ell^3 \left(6 c_4 \ell + 12 \ell q_1{}^2 + 3 c_3 r + 2 q_1 r v_2 - \ell v_2{}^2 + 3 \ell q_1 v_3\right)}{r^6} \tag{D2}$$

$$C^{\hat{4}}{}_{\hat{4}} = \frac{\ell^3 \left(-3 c_4 \ell - 6 \ell q_1{}^2 - 2 c_3 r - 2 q_1 r v_2 + \ell v_2{}^2 - 3 \ell q_1 v_3\right)}{r^6} \tag{D3}$$

$$C^{\hat{4}}{}_{\hat{5}} = \frac{\ell^2 v_2}{r^4} + \frac{\ell^3 \left(-(m_1 v_2) + 3 v_3\right)}{r^5}$$
$$+ \frac{\ell^4 \left(-48 m_1 q_1 - m_1{}^2 v_2 + q_1{}^2 v_2 - 6 m_1 v_3 + 12 v_4\right)}{2 r^6}. \tag{D4}$$

The field equations (B1) and the lowest order terms of Eq. (D4) lead immediatly to $v_2 = v_3 = 0$. Using this result in Eq. (D3), one finds that also $c_3 = 0$. With

$$v_2 = v_3 = c_3 = 0, \tag{D5}$$

the left-hand side of the field equations (to fourth order in $\ell$) takes the simple form

$$C^{\hat{1}}{}_{\hat{1}} = \frac{-3 \ell^4 \left(c_4 + 2 q_1{}^2\right)}{r^6}$$

$$C^{\hat{2}}{}_{\hat{2}} = \frac{3 \ell^4 \left(2 c_4 + 4 q_1{}^2\right)}{r^6}$$

$$C^{\hat{4}}{}_{\hat{4}} = \frac{-3 \ell^4 \left(c_4 + 2 q_1{}^2\right)}{r^6}$$

$$C^{\hat{4}}{}_{\hat{5}} = \frac{6 \ell^4 \left(-4 m_1 q_1 + v_4\right)}{r^6}.$$

These equations imply

$$c_4 = -2 q_1^2 \quad \text{and} \quad v_4 = 4 m_1 q_1. \tag{D6}$$

# APPENDIX E: SHORT DISTANCE BEHAVIOUR

Assume that the short-distance behaviour can be described with a power-law solution. From Eqs. (C6)–(C8) we see that for this to be possible $L(r)$ and $a(r)$ must be related as

$$a(r) = \frac{1}{2} \ln \left[\beta^{-2} L(r)^{-2}\right] \tag{E1}$$

and

$$a'(r) = \frac{k}{r} \tag{E2}$$

where both $\beta$ and $k$ are dimensionless constants. We also have to neglect the terms proportional to $\ell^{-2}$, in accordance with restricting the validity of the solution to the region where $r \ll \ell$. Using Eqs. (E1) and (E2) to compute the derivatives, substituting back into Eqs. (C6–C8), and taking the limit $\ell \to \infty$, we find that the only non-vanishing component of the field equations is

$$C^{\hat{1}}_{\hat{1}} = \frac{-1 + \beta^2 + k - \beta^2 k}{r^2}. \tag{E3}$$

The solution of $C^{\hat{1}}_{\hat{1}} = 0$ is $k = 1$.

The short-distance power-law solution is

$$e^{2a(r)} = \frac{r^2}{r_0^2}, \quad L(r) = \beta \frac{r_0}{r} \quad \text{and} \quad V(r) = V_0 \pm \frac{r^2}{2\ell r_0}. \tag{E4}$$

### APPENDIX F: FOUR-DIMENSIONAL ENERGY-MOMENTUM TENSOR

The five-dimensional field equations $C^{\mu}_{\nu} = 0$ can be split into three parts in four-dimensional space-time. Some terms correspond to a four-dimensional Einstein tensor. The remaining terms can be interpreted as the electromagnetic energy-momentum tensor. The latter can be separated in two parts, one coming from the five-dimensional Einstein-Hilbert term (standard Maxwell terms), and another part coming from the Gauß-Bonnet term. Thus we write

$$C^i_j = {}^{(4)}G^i_j - \kappa T(M)^i_j - \kappa T(G_B)^i_j = 0. \tag{F1}$$

Using the expressions of Appendix B, we get

$$\kappa T(M)^i_j = \frac{-V'(r)^2}{4\, e^{2a(r)}} \, \text{diag}[1, -1, -1, 1]. \tag{F2}$$

This is the Maxwell energy-momentum tensor. It is manifestly traceless and radially boost invariant.

The Gauß-Bonnet term gives rise to the following energy-momentum tensor

$$\kappa T(G_B)^r_{\ r} = \frac{\ell^2 \left(3 - e^{2a(r)} L(r)^2\right) V'(r)^2}{e^{4a(r)} r^2 L(r)^2} \tag{F3}$$

$$\kappa T(G_B)^{\theta}_{\ \theta} = \frac{\ell^2 V'(r) \left(-3 L(r) a'(r) V'(r) - 2 L'(r) V'(r) + 2 L(r) V''(r)\right)}{e^{4a(r)} r L(r)^3} \tag{F4}$$

$$\kappa T(G_B)^t_{\ t} = \frac{\ell^2 \left(1 - e^{2a(r)} L(r)^2\right) V'(r)^2}{e^{4a(r)} r^2 L(r)^2} \tag{F5}$$

This energy-momentum tensor violates radial boost invariance because

$$\kappa T(G_B)^r_{\ r} - \kappa T(G_B)^t_{\ t} = \frac{2\,\ell^2 V'(r)^2}{e^{4a(r)} r^2 L(r)^2}. \tag{F6}$$

The trace of the energy-momentum tensor coming from the Gauß-Bonnet term is

$$\kappa T(G_B)^i_i = 2\ell^2 V'(r) \left[ 2 L(r) V'(r) - e^{2a(r)} L(r)^3 V'(r) - 3 r L(r) a'(r) V'(r) \right.$$
$$\left. - 2 r L'(r) V'(r) + 2 r L(r) V''(r) \right] \left[ e^{4a(r)} r^2 L(r)^3 \right]^{-1}. \tag{F7}$$

By substituting the solutions of Appendix C into the above expression, we get

$$\kappa T(G_B)^i_i = \left[ \ell^2 V'(r)^2 \left( 48 \ell^2 - 48 e^{2a(r)} \ell^2 L(r)^2 + 12 e^{2a(r)} r^2 L(r)^2 - 16 \ell^4 L(r)^2 V'(r)^2 \right. \right.$$
$$- 4 \ell^2 r^2 L(r)^2 V'(r)^2 + 16 e^{2a(r)} \ell^4 L(r)^4 V'(r)^2 + 8 e^{2a(r)} \ell^2 r^2 L(r)^4 V'(r)^2$$
$$\left. + e^{2a(r)} r^4 L(r)^4 V'(r)^2 \right) \right] \left[ 2 e^{4a(r)} r^2 L(r)^2 \left( 4\ell^2 - 4 e^{2a(r)} \ell^2 L(r)^2 \right. \right.$$
$$\left. \left. - e^{2a(r)} r^2 L(r)^2 \right) \right]^{-1}. \tag{F8}$$

Thus, the trace of the energy-momentum tensor does not vanish in general.

## APPENDIX G: THE TWO-BODY CENTRAL FORCE PROBLEM AND (2+1)-DIMENSIONAL COSMOLOGY

The non-relativistic two-body central force problem in three dimensional space, reduces to an effectively two-dimensional problem because the central force motion is always in a plane. By using the center of mass frame, it can further be reduced to an equivalent one-body problem. It can be proved that the only potentials which produce closed orbits for all bound particles are the $1/r$ and $r^2$ potentials, i.e. Coulomb-type interaction and Hooke's law. Why these potentials are singled out is not so evident. Here we show that the central force problem may be mapped onto a (2+1)-dimensional homogeneous and isotropic cosmological model, and that these special force laws correspond to a traceless energy-momentum tensor and a vacuum energy-momentum tensor ($\Lambda$-term), respectively.

The first integral in the center of mass system is given as

$$m\dot{r}^2 = 2K - 2U(r) \tag{G1}$$

where $K$ is the energy and $U(r)$ is the effective potential (including the centrifugal terms if the angular momentum is non-zero). Solutions of this equation correspond to solutions of Einstein's equations

$$R_{\mu\nu} - \frac{1}{2} g_{\mu\nu} R = \frac{1}{m} T_{\mu\nu} \tag{G2}$$

for a (2+1)-dimensional Friedmann-Lemaître-Robertson-Walker cosmology. With the line-element

$$ds^2 = a(t)^2 \left( \frac{1}{1-k\rho^2} d\rho^2 + \rho^2 d\theta^2 \right) - dt^2 \tag{G3}$$

the Einstein tensor is

$$G^\rho_{\ \rho} = G^\theta_{\ \theta} = -\frac{\ddot{a}}{a}, \quad G^t_{\ t} = -\frac{k}{a^2} - \frac{\dot{a}^2}{a^2}. \tag{G4}$$

Thus, the (2+1)-dimensional FLRW cosmology is in one-to-one correspondence with our problem provided the following identifications are made

$$a(t) = r(t), \quad k = -2K/m, \quad \varrho = -2U(r)/r^2, \quad p = U'(r)/r \tag{G5}$$

where $a(t)$ is the cosmic scale factor, $k$ is the spatial curvature, $\varrho$ is the energy-density of a perfect fluid, and $p$ is the pressure of the fluid. This energy-momentum tensor is covariantly conserved for all effective potentials $U(r)$. For vanishing angular momentum, one readily finds that a $1/r$ potential leads to a $\rho \propto 1/a^3$, corresponding to an ultra-relativistic fluid, and that a $r^2$ potential leads to $\rho$ constant, corresponding to a $\Lambda$-term.

# REFERENCES

[1] Th. Kaluza, *Sitzungsber. Preuss. Akad. Wiss. Leipzig* (1921), 966.
[2] O. Klein, *Z. Phys.* **37** (1926), 895.
[3] Ø. Grøn, *Il Nuovo Cimento* **91** B (1986), 57.
[4] Ø. Grøn and P. Ødegaard, *Gen. Rel. Gravit.* **26** (1994), 53.
[5] O. Klein, *Nature* **118** (1926), 516.
[6] A. J. Schwarz and N. A. Doughty, *Am. J. Phys.* **60** (1992), 150.
[7] R. Kerner, *Ann. Inst. H. Poincaré* **9** (1968), 143.
[8] A. Salam and J. Strahdee, *Ann. Phys. (N. Y.)* **141**, (1982), 316.
[9] Further references are found *in* "Modern Kaluza-Klein Theories," (T. Appelquist, A. Chodos, and P. G. O. Freund, Eds.), Addison Wesley Publ. Co., 1987.
[10] Y. Thiry, *Comptes Rendus* **226** (1948), 216.
[11] E. Leibowitz and N. Rosen, *Gen. Rel. Gravit.* **4** (1973), 449.
[12] D. Lovelock, *J. Math. Phys.* **12** (1971), 498.
[13] B. Zwiebach, *Phys. Lett. B* **156** (1985), 315.
[14] B. Zumino, *Phys. Rep.* **137** (1986), 109.
[15] J. T. Wheeler, *Nucl. Phys. B* **273** (1986), 732.
[16] J. Scherk and J. H. Schwarz, *Nucl. Phys. B* **81** (1974), 118.
[17] S. Deser and A. N. Redlich, *Phys. Lett. B* **176** (1986), 350.
[18] D. Hochberg and T. Shimada, *Progr. Theor. Phys.* **78** (1987), 680.
[19] R. R. Metsaev and A. A. Tseytlin, *Nucl. Phys. B* **293** (1987), 385.
[20] D. Lüst, S. Theisen, and G. Zoupanos, *Nucl. Phys. B* **296** (1988), 800.
[21] C. Aragone, *Phys. Lett. B* **186** (1987), 151.
[22] I. Jack and D. R. T. Jones, *Nucl. Phys. B* **303** (1988), 260.
[23] S. Deser, *in* "Proceedings of the 2nd Canadian Conference on General Relativity and Relativistic Astrophysics," (A. Coley, C. Dyer, and T. Tupper, Eds.), p. 225, World Scientific, Singapore, 1988.
[24] S. Kalara, C. Kounnas, and K. A. Olive, *Phys. Lett. B* **215** (1988), 265.
[25] V. A. Kostelecký and S. Samuel, *Phys. Rev. Lett.* **63** (1989), 224.
[26] B. A. Campbell, M. J. Duncan, N. Kaloper, and K. A. Olive, *Nucl. Phys. B* **351** (1991), 778.
[27] H. A. Buchdahl, *J. Phys. A: Math. Gen.* **12** (1979), 1037.
[28] R. Kerner, *C. R. Acad. Sc. Paris, série II* **304** (1987), 621.
[29] F. Müller-Hoissen, *Phys. Lett. B* **201** (1988), 325.
[30] G. W. Horndeski, *J. Math. Phys.* **17** (1976), 1980.
[31] D. G. Boulware and S. Deser, *Phys. Rev. Lett.* **55** (1985), 2656.
[32] G. W. Gibbons and P. J. Ruback, *Phys. Lett. B* **171** (1986), 390.
[33] A. Tomimatsu and H. Ishihara, *J. Math. Phys.* **28** (1987), 2720.
[34] J. T. Wheeler, *Nucl. Phys. B* **268** (1986), 737.
[35] D. L. Wiltshire, *Phys. Lett. B* **169** (1986), 36.
[36] D. L. Wiltshire, *Phys. Rev. D* **38** (1988), 2445.
[37] E. Poisson, *Phys. Rev. D* **43** (1991), 3923.
[38] B. Whitt, *Phys. Rev. D* **38** (1988), 3000.
[39] J. Madore, *Phys. Lett. A* **110** (1985), 289.
[40] J. Madore, *Phys. Lett. A* **111** (1985), 283.

[41] K. Maeda, *Phys. Lett. B* **166** (1986), 59.
[42] N. Deruelle and J. Madore, *Phys. Lett. A* **114** (1986), 185.
[43] N. Deruelle and J. Madore, *Mod. Phys. Lett. A* **1** (1986), 237.
[44] H. Ishihara, *Phys. Lett. B* **179** (1986), 217.
[45] F. Müller-Hoissen, *Class. Quant. Grav.* **3** (1986), 665.
[46] A. B. Henriques, *Nucl. Phys. B* **277** (1986), 621.
[47] N. Deruelle and J. Madore, *Phys. Lett. B* **186** (1987), 25.
[48] B. Giorgini and R. Kerner, *Class. Quant. Gravit.* **3** (1988), 339.
[49] P. Halpern and D. Kerrick, *Gen. Rel. Gravit.* **23** (1993), 41.
[50] G. Nordström, *Phys. Zeitschr.* **15** (1914), 504.
[51] C. W. Misner, K. S. Thorne, and J. A. Wheeler, "Gravitation," W. H. Freeman, San Francisco, 1973.
[52] C. Itzykson and J.-B. Zuber, "Quantum Field Theory," McGraw-Hill, New York, 1985.
[53] W. Heisenberg and H. Euler, *Zeitschr. Phys.* **98** (1936), 714.
[54] I. T. Drummond and S. J. Hathrell, *Phys. Rev. D* **22** (1980), 343.
[55] M. Novello, L. A. R. Oliveira, and J. M. Salim, *Class. Quant. Gravit.* **7** (1990), 51.
[56] N. D. Birrell and P. C. W. Davies, "Quantum Fields in Curved Space," Cambridge University Press, Cambridge, 1982.
[57] S. L. Adler, *Ann. Phys. (N. Y.)* **67** (1971), 599.
[58] E. Iacopini and E. Zavattini, *Phys. Lett. B* **85** (1979), 151.

FIGURES

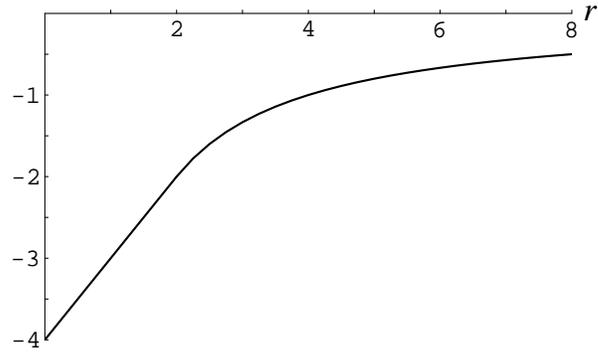

FIG. 1. The effective modified Coulomb potential $V_{\text{eff}} = -q\varphi(r)$ seen by a particle with charge $q = 2\ell$ in the field of another particle with the charge $Q = -2\ell$. Mass-dependent corrections have been neglected. Both $V_{\text{eff}}$ and the radial distance are given in units of $\ell$.